\documentclass[letterpaper]{aa}
\usepackage{graphics}
\usepackage{txfonts}
\usepackage{natbib}

\def\xiunit{\,erg\,cm\,s$^{-1}$}
\begin{document}

\title
{
The variable X-ray spectrum of Markarian 766 - I.\\
Principal components analysis
}
\subtitle{}

\titlerunning{Principal components analysis of Mrk\,766}

\author
{L.\ Miller\inst{1} \and 
T.\ J.\ Turner\inst{2,3} \and 
J.\ N.\ Reeves\inst{3,4} \and 
I.\ M.\ George\inst{2,3} \and
S.\ B.\ Kraemer\inst{5,3} \and
B.\ Wingert\inst{2}
}

\authorrunning{L.\ Miller et al.\ }

\institute{Dept. of Physics, University of Oxford, 
Denys Wilkinson Building, Keble Road, Oxford OX1 3RH, U.K.
\and 
Dept. of Physics, University 
of Maryland Baltimore County, 1000 Hilltop Circle, Baltimore, MD 21250, U.S.A.
\and 
X-Ray Astrophysics Laboratory, 
Code 662,  Astrophysics Science Division,   
NASA/GSFC, Greenbelt, MD 20771, U.S.A.
\and
Dept. of Physics and Astronomy, Johns Hopkins University, 3400 N 
Charles Street, Baltimore, MD 21218, U.S.A.
\and
Catholic University of America, Washington DC 20064, U.S.A.
}

\date{Received / Accepted}

\abstract
{}
{We analyse a long XMM-Newton spectrum of the narrow-line Seyfert\,1
galaxy Mrk\,766, using the marked spectral variability on timescales
$>20$\,ks to separate components in the X-ray spectrum.
}  
{Principal components analysis is used to identify
distinct emission components in the X-ray spectrum, possible alternative
physical models for those components are then compared statistically.
}
{
The source spectral variability is well-explained by additive
variations, with smaller extra contributions most likely arising from
variable absorption.  The principal varying component, eigenvector
one, is found to have a steep (photon index 2.4) power-law shape,
affected by a low column of ionised absorption that leads to the
appearance of a soft excess.  Eigenvector one varies by a factor 10 in
amplitude on time-scales of days and appears to have broad ionised
Fe\,K$\alpha$ emission associated with it: the width of the ionised
line is consistent with an origin at $\sim 100$ gravitational radii.
There is also a strong component of near-constant emission that
dominates in the low state, whose spectrum is extremely hard above
1\,keV, with a soft excess at lower energies, and with a strong edge
at Fe\,K but remarkably little Fe\,K$\alpha$ emission.  Although this
component may be explained as relativistically-blurred 
reflection from the inner accretion disc, 
we suggest that its spectrum and lack of variability may
alternatively be explained as either (i) ionised reflection from an extended
region, possibly a disc wind, or (ii) a signature of absorption by a
disc wind with a variable covering fraction.  Absorption features in
the low state may indicate the presence of an outflow.
}
{}


\maketitle

\section{Introduction}

X-ray observations of many AGN show substantial flux variations, with
accompanying systematic spectral variations.  It is likely that the
X-ray emission from AGN is composed of a number of emission
components, which may or may not be interrelated, modified by the
effects of absorption. The basic components of the AGN model are 
thought to be a 
continuum power-law that reflects off the surface of the accretion disk
\citep{guilbertrees88,lightmanwhite88,georgefabian} 
producing an observable spectral 
hardening above 10 keV   \citep{zdziarski95,perola02} 
and strong Fe K shell emission (e.g. \citealt{tanaka95}, 
\citealt{nandraea97}). Layers of gas 
are also thought to shroud the nuclear system, with a large range of 
ionisation-states and 
column densities \citep{crenshaw03}. 

While emission, reflection and absorption are notoriously difficult 
to disentangle in the mean X-ray spectra of AGN \citep{reeves04,turnerea05}, 
observed spectral changes over time can be used to attempt to
decompose the emission into physically-distinct components 
\citep{taylor, uttley, vaughanfabian04}. By correlating
the flux measured at different energies it is possible to determine
the spectrum of two components: one assumed to be a constant
zero-point spectrum and the other to be a component whose amplitude
varies with time.
The conclusion from the \citet{vaughanfabian04} study of MCG--6-30-15
has been that there exists an underlying component of X-ray emission 
with a hard spectrum that appears consistent with that expected
from reflection of the X-ray continuum by optically-thick gas.  

Here we present results of statistical analysis of the X-ray spectral
variability of Mrk~766 using principal components analysis.
Mrk~766
has one of the largest integrated exposure times to date from {\it
XMM-Newton}. The large amount of accumulated data combined with the
high degree of flux variability exhibited in the X-ray band makes
Mrk~766 an ideal target for this approach.  Mrk\,766 is a narrow-line
Seyfert\,1 at redshift $z=0.0129$ \citep{osterbrock}.  
It is known from earlier observations to have a broad component of 
Fe\,K$\alpha$ emission \citep{pounds03a}. \citet{turner766} showed
that there appear to be variations in line energy consistent with 
an origin at around 100 gravitational radii, r$_g$, 
and, using the same data analysed in this paper,
\citet{miller06}
showed that the ionised component of the line varies in flux with the
continuum, implying an origin for the line emission consistent with
that estimate.  The full dataset discussed here shows that Mrk\,766 exhibits a very
wide range of flux levels, with marked spectral variability, and in this
paper we attempt to use that information to learn more about the emission
and absorption components.  The variation timescales used are $> 20$\,ks,
complementing the analysis of rapid variability by \citet{markowitz}.
In Paper\,II (Turner et al., in preparation) we shall investigate direct model
fits to the full Mrk\,766 dataset, following up on the general behaviour
discovered here.

In section\,2 we summarise the method of principal components analysis and describe
briefly its limitations.  We have adopted a mathematical decomposition that allows us
to extract component spectra with high spectral resolution despite a limit imposed
by having a finite amount of data.  The data used are described in section\,3, and
section\,4 describes the principal components analysis and a statistical analysis
of the results:  a key point here is to decide whether or not the principal
components analysis provides a good description of the source variations.  Section\,5
then fits some possible alternative models to the spectra of the derived components,
and section\,6 discusses the physical properties of the source that would be required
in each model.  

\section{Principal Components Analysis}
\subsection{Introduction}
\label{pca_introduction}
A powerful method of decomposing time-variable data is
to use Principal Components Analysis
(hereafter PCA).  PCA is widely used in multivariate analysis and has
previously been used to understand the differing components of
emission that comprise the optical spectra of samples of
QSOs \citep{francis,boroson} and of multiple observations of individual
active galaxies \citep{mittaz}, 
and has recently been applied to the X-ray
spectra of Seyfert galaxy MCG--6-30-15 by \citet{vaughanfabian04}.
The mathematical model adopted is essentially the same
as that assumed in the flux-flux correlation
method: namely that there exist 
spectrally-invariant components of emission whose amplitude may vary
with time.  These components can then be considered as being linearly
combined to produce the observed spectra.  PCA allows multiple components
to be detected.  In principle any one
multivariate dataset may be decomposed into as many components as
there are measurements: as we shall describe below, a powerful
advantage of using PCA is that we can test how many such components
are required to explain the observed variations.  The model of
spectrally-invariant components may be violated in real AGN emission:
an obvious example would be if there were time-varying absorption,
which cannot be modelled as a
series of additive components.  Power-law components may also vary
their spectral indices \citep{uttley}.  However, we may test whether
such effects are present by investigating the required number of
components and their shape.

To visualise the application of PCA to time-varying spectra, consider
a spectrum made up of $n$ measurements, at different times, of the
flux in $m$ bins of photon energy.  Consider an $m$-dimensional space
in which each axis is the flux in each photon energy bin: one of the
measured spectra then is described by a single point in that
$m$-dimensional space.  The full dataset comprises a distribution of
$n$ such points.  If there were no variations all the points would be
co-located, but if there were a single component of varying emission
the distribution of points would be dispersed along a straight line
passing through the origin.  In general, if there were $p$ such
components, the cloud of points would be confined to a $p$-dimensional
surface embedded in the $m$-dimensional space.  If one of those
components in fact were constant, the $p$-dimensional surface could
also be described in $p-1$ dimensions with a shift of the origin.  One
of the tasks we shall carry out is to evaluate how many dimensions are
required to explain the data variations, and whether the variations
may be equally well described by having one component constant.

The PCA consists of finding a coordinate rotation of the
$m$-dimensional space so that one axis in the rotated system lies in
the direction for which the distribution has the largest variance.
This direction is known as the first principal component, or eigenvector one.  
The second
principal component is orthogonal to the first and is the axis along
which the distribution has the next largest variance, and so on.  It
is conventional to first shift the coordinate origin to the mean of
the data distribution - this is a convenient choice as it effectively
removes any constant component and allows such data to be described by
$p-1$ components: the presence of a constant component can be
determined by establishing whether or not the $p-1$-dimensional
surface passes through the original origin.  Mathematically, the
principal components are the eigenvectors of the data covariance
matrix, and the variances along each component are the eigenvalues.
In our case of analysing spectra, each eigenvector is the spectrum of
an individually varying component, and the eigenvalue measures how
much variation that component is responsible for.

\begin{figure}
\begin{minipage}{70mm}{
\resizebox{70mm}{!}{
\rotatebox{0}{
\includegraphics{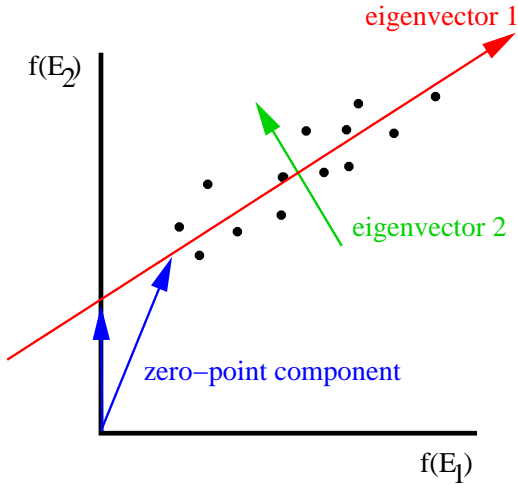}}}}
\end{minipage}
\caption{
Schematic explanation of the relationship between PCA eigenvectors
and the zero-point spectrum.  The axes correspond to the flux measurements
in each spectral bin (only two shown here), each point corresponds
to observations in a different timeslice.  The eigenvectors are orthogonal
components that span the plane, the zero-point spectrum may be constructed
by a linear combination of the mean spectrum and the eigenvectors, and is
therefore in general neither orthogonal to the eigenvectors nor unique.
The minimum and maximum possible values of the zero-point spectrum 
are indicated by the two arrows that start at the origin.
}
\label{pcademo}
\end{figure}

The principal components are orthogonal, in the sense that in the
rotated coordinate system there is no covariance between the
components.  However, these orthogonal components need not correspond
to physical components, since there need be no requirement for the
true physical components to vary independently.  Thus in reaching a
physical interpretation of the analysis, we may need to allow that the
physical components may be non-orthogonal and may be constructed out
of arbitrary linear combinations of the principal components.  Even in
the case where the emission may be modelled as a constant zero-point
spectrum plus a single varying component, the zero-point is not
uniquely defined as it may lie anywhere along the vector defined by
the first principal component.  If we assume that both the first
principal component and the zero-point spectrum are positive
(corresponding to being emission components), however, then we may
place limits on possible values of the zero-point spectrum.  A lower
bound is given by requiring that no photon energy bin should have a
negative value; an upper bound is given by requiring that no observed
spectrum should lie below the zero-point spectrum (as this would
require a negative contribution from the varying component).  Arbitrary
spectra between these two extremes are not allowed: the zero-point
spectrum must lie along the vector defined by the varying component.
Note that this zero-point uncertainty is also present in the flux-flux
correlation method described above.  
Fig.\,\ref{pcademo} illustrates the relationship between these various
components.

One question to be investigated in the data is whether such a constant
zero-point is required by the data, or whether the data are instead
consistent with all spectral components being required to vary between
timeslices.  We shall approach this by doing two principal components
analyses.  In the first, we shall allow there to be a zero-point spectrum
as described.  In the second, we shall calculate the covariance matrix
about the flux origin rather than the data mean, which in effect forces
all spectral components to be variable.  

\subsection{Singular Value Decomposition}
A particular problem arises if the number of datapoints $n$ is less than
the number of dimensions $m$.  This situation arises in the analysis of
the {\em XMM-Newton} data described below, where in order to obtain adequate
signal-to-noise in each timeslice the data are limited to 
$n \la 60$ timeslices, but where the spectra are binned into $m \simeq 140$
energy intervals.  In this case the covariance matrix is singular and the
full set of eigenvectors needed to completely describe the data is not 
uniquely defined.  However, the least significant
eigenvectors simply describe shot noise in the data, whereas the
most significant eigenvectors remain well-defined, being the principal 
axes along which the source varies.  These leading eigenvectors may be
extracted from the singular covariance matrix using Singular Value
Decomposition (e.g. \citealt{numrecipes}) and this is the technique
used here.  Hence the component spectra are reproduced at the full
instrumental resolution of {\em XMM-Newton}, unlike previous analyses
of X-ray data \citep{vaughanfabian04}.

\subsection{The effects of absorption}
Finally, we should return again to the question of absorption.  If any
component suffers time-invariant absorption, then this will not
prejudice the analysis, it will simply show up as a modulation of the
derived component spectra.  Only in the case where the absorption
varies with time will additional components apparently be created.
We address this here by first considering PCA only of data in the energy
range $2-10$\,keV where the effects of absorption should be smallest,
and then seeing the effect of extending the PCA to lower energies.

\section{The data}

In this paper we utilise all {\it XMM-Newton} \citep{jansen} EPIC pn 
\citep{struder} data available
for Mrk~766.  We reanalysed the archival observations from 2000 May 20,
science observation ID 0096020101 \citep{boller}, 
and from 2001 May 20-21, science observation ID 0109141301 \citep{pounds03a}.  
We also analyse
a long observation made during 2005 May 23 UT 19:21:51 -
Jun 3 UT 21:27:10 over six {\it XMM-Newton} orbits, science observation IDs
in the range 0304030[1-7]01.  The same observations were also analysed
by \citet{miller06}.  EPIC pn data were screened in the standard way
using SAS v6.5 software to select only events with patterns in the
range 0 - 4. SAS v6.5 was also used to generate response matrices. We
applied energy cuts to the events files to discard data below 0.2 keV
and above 15.0 keV.  Background filtering then removed any periods
where the count rate in the background cell 
(a source-free region of area 3\,arcmin$^2$ on the same chip as the
target) exceeded 2 counts\,s$^{-1}$, and any periods where the
background rate exceeded 5\% of the source count rate. This combined
filtering yielded 24\,ks of `good' data for 2000 May, 67\,ks for 2001
May and 402\,ks of `good' data over a 944\,ks baseline during 2005
May-June.  Target photons were extracted from a circular region of $40''$
radius centred on the Mrk~766.  The combination of some non-optimal
MOS modes, photon pileup and inferior signal-to-noise led us to use
only the pn data in this analysis.

\section{Principal components analysis of Mrk\,766}
\label{sec:pca}
\subsection{Generation of time-sliced spectra and the principal components}
\label{sec:generation}
To carry out the PCA we must first divide the data into timesliced
spectra that are binned in energy.  
The energy bins cannot be infinitesimally small because shot noise
would dominate over intrinsic source variations, and the PCA would
be meaningless.  In order to maximise the spectral information,
we choose energy bins equal in width to the half-width at half-maximum (HWHM) 
of the EPIC pn
instrument.  The HWHM varies with photon energy, so the bin widths 
chosen also vary with energy, from $\sim 35$\,eV at $E \sim 1$\,keV
to $\sim 80$\,eV at $E \sim 10$\,keV.  Hence the spectra have effectively been
smoothed with a function approximately equal in width to half the 
energy-dependent instrumental energy resolution, and sampled at that interval.  
The statistical analysis of the principal components discussed in 
sections\,\ref{sec:2-10}\,\&\,\ref{sec:0p4} and the model-fitting discussed in 
section\,\ref{modelling} are all carried out on data analysed in this way, with
spectral bins whose photon shot noise is statistically independent.  To
enable better visual detection of features in the spectra, the spectra displayed
in sections\,\ref{sec:PCspectra}\,\&\,\ref{sec:datacomparison} 
have also been smoothed with a top-hat of width two bins:
in this way the spectra have been smoothed with a function close to the width of
the instrumental resolution, improving the detectability of weak features
but introducing greater covariance between spectral data points and degrading
the resolution by a factor $\sqrt{2}$ compared with the instrumental resolution.

The signal-to-noise
in each timesliced energy bin increases with the source brightness and
the duration of each time slice.  The time slicing must also not be
chosen to be so coarse that it washes out the temporal spectrum variations.
For Mrk\,766 time slicing of 20\,ks was found to provide a good compromise
between these two criteria.  Even so, the signal-to-noise at high
energies, $E > 7.5$\,keV was found to be low, and these energy bins
were grouped together until their signal-to-noise, averaged across all the
observations, reached a value of 10 in each bin - 
degrading the spectral resolution at high energies
but ensuring that the PCA remains meaningful.  

The energy range chosen for the analysis initially is $2-9.8$\,keV, 
although in section\,\ref{sec:0p4}
we will extend the range to lower energies, $0.4 < E < 9.8$\,keV.
In both ranges the effective area of the
EPIC pn detector exceeds $\sim 250$\,cm$^2$.
For brevity, in the remainder of the paper we shall refer to these
energy ranges as $2-10$ and $0.4-10$\,keV respectively. 
The spectra are 
analysed and displayed in units of Ef(E), where f(E) is the spectral
flux density.  This choice means that a power-law with photon index
$\Gamma = 2$ has a uniform value with energy and thus gives
approximately equal weight to each photon energy bin in the spectra
considered here.  As the principal components are additive, 
each timeslice is given a weight inversely proportional
to its flux so that the fractional spectral variations at low flux levels
are given equal importance to those at high flux levels: otherwise the
PCA results would be dominated by the high flux states, whereas we are
seeking a principal components model that gives a good description of
the entire dataset.  With this weighting scheme, if we reconstruct a model 
using a subset of the principal components, we expect the data/model ratio
in different timeslices to have deviations of similar amplitude 
irrespective of flux.

\subsection{Error analysis}\label{sec:errors}
Errors on the resulting component spectra are obtained here by a 
Monte-Carlo method, in which the observed photon counts binned in
energy and time are perturbed by a random amount commensurate with
the photon shot noise and the PCA redone on the perturbed dataset.
This process is repeated 20 times to make an estimate of the variance
on the component spectra at each binned energy.  Note that the errors
are correlated between the various components - a deficit owing to noise
on one component spectrum will be accompanied by an excess on another
component at the same energy. 

\subsection{Results: $2-10$\,keV}\label{sec:2-10}
We first analyse spectral variations in the range $2-10$\,keV, as this 
range should be relatively immune to the non-additive effects of any
varying absorption.  We first carry out the PCA assuming
the existence of a constant zero-point spectrum.

\begin{figure}
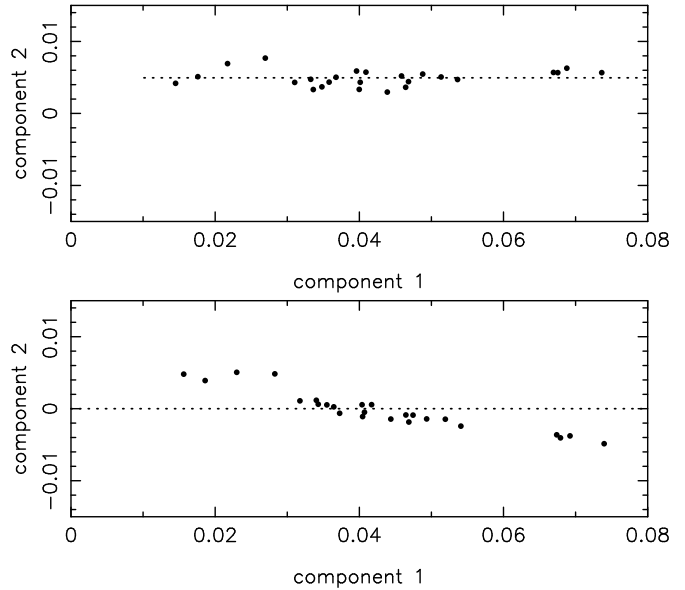


\begin{minipage}{88mm}{
\resizebox{88mm}{!}{
\rotatebox{270}{
\includegraphics{6548fig2a.ps}}}}
\end{minipage}

\begin{minipage}{88mm}{
\resizebox{88mm}{!}{
\rotatebox{270}{
\includegraphics{6548fig2b.ps}}}}
\end{minipage}

\caption{
{\em Top:}
Amplitudes of the first two principal components for Mrk\,766
assuming the existence of a constant zero-point spectrum
for each individual time sliced spectrum:
1st component (x-axis) v. 2nd component (y-axis).
The origin is defined 
as the location of a spectrum with zero flux.  The 1st
principal component vector is shown as the dashed line.  The
lower bound to this line defines the point below which the
spectrum would have negative portions if composed of these two components.
{\em Bottom:}
As above, but with no zero-point spectrum.
}
\label{mkn766_2-10keV_comp1_comp2}
\end{figure}

\begin{figure}
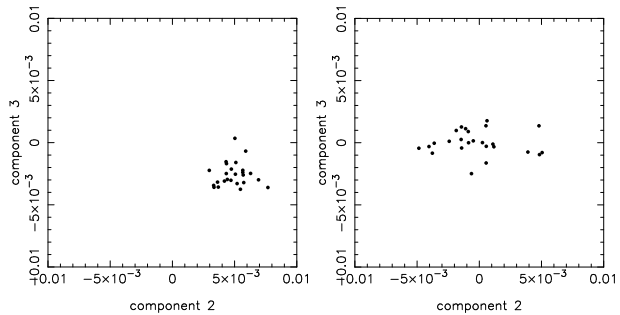

\begin{minipage}{40mm}{
\resizebox{40mm}{!}{
\rotatebox{270}{
\includegraphics{6548fig3a.ps}}}}
\end{minipage}
\begin{minipage}{40mm}{
\resizebox{40mm}{!}{
\rotatebox{270}{
\includegraphics{6548fig3b.ps}}}}
\end{minipage}

\caption{
{\em Left:}
A rotated view of the 2nd and 3rd principal components for Mrk\,766, 
looking along the axis of the 1st principal component, and showing the
offset of the distribution from the origin assuming there is a 
constant zero-point spectrum.
{\em Right:}
As above, but with no zero-point spectrum.
}
\label{mkn766_2-10keV_comp2_comp3}
\end{figure}

We find that a reasonable description of the data can be found allowing
a zero-point spectrum and a single varying component (eigenvector one).
The upper panel of
Fig.\ref{mkn766_2-10keV_comp1_comp2} shows the
amplitudes of the first two components shifted so that a spectrum
with zero flux would lie at the location (0,0) on this graph.  The 1st
component plotted on the x-axis has a range of values 10 times larger
than the second component, and, as described below, the first
component accounts for $96$\,percent of the variance about the mean spectrum.  
The extrapolation of this component does not pass
through the flux origin and hence an offset zero-point
spectrum is required.  
This is also clear if we rotate our view to look along the
1st component, and plot the scatter about that axis in terms of the
next two principal components 
(left panel of Fig.\,\ref{mkn766_2-10keV_comp2_comp3}):
there is substantially less scatter in these two components than in
the first component, and the data distribution sits away from the flux
origin.  Table\,\ref{mkn766_2-10table} gives an indication of the number of
components required to describe the data.  The first column identifies the
component, these being ordered by their eigenvalues.
The second column gives the
eigenvalue associated with each component, which is equivalent to the
variance along that component.  The third column
gives the proportion of the total variance that is described by this
component.  These two measures give an indication of how much of the
observed variation is explained by each component.  It is clear that
the first component dominates the variation.

\begin{table}
{\small
\begin{tabular}{lr@{.}ll@{}rr@{.}ll@{}r}
& \multicolumn{4}{c}{with zero-point spectrum} & \multicolumn{4}{c}{without zero-point spectrum}\\
\# & \multicolumn{2}{c}{$\lambda$} &
$\sigma_f^2$ & $\chi^2$/dof & \multicolumn{2}{c}{$\lambda$} & $\sigma_f^2$ & $\chi^2$/dof \\
& \multicolumn{2}{c}{$/10^{-5}$} & & & \multicolumn{2}{c}{$/10^{-5}$} & & \\
0 &  &   & 0.     & 4993/106 &    &  & 0.     & 23961/106 \\
1 & 22&5 & 0.956  & 122/105  & 151&1 & 0.988  & 371/105 \\
2 & 0&14 & 0.0059 & 96/104  & 0&94  & 0.0061 & 99/104 \\
3 & 0&092 & 0.0039 & 87/103  & 0&11  & 0.0007 & 91/103 \\
4 & 0&082 & 0.0035 & 80/102  & 0&086  & 0.0006 & 83/102 \\
5 & 0&077 & 0.0033 & 73/101   & 0&077  & 0.0005 & 76/101 \\
\end{tabular}
}
\caption{
The amount of spectral variation explained by the first five 
principal components for Mrk\,766 in the energy range $2-10$\,keV.
Two cases are shown, one assuming there is a constant zero-point spectrum,
the other without.  Columns $2-4$ and $5-6$ give the eigenvalue $\lambda$,
fractional variance $\sigma_f^2$, $\chi^2$ with numbers of degrees of
freedom (dof) for each case.
The fractional variances are not comparable
between these two cases as the variances are defined differently.
}
\label{mkn766_2-10table}
\end{table}

This does not mean of course that the remaining components are not
significant.  We can also attempt to evaluate how well the data are
described by a model comprising a zero-point spectrum plus some number
of components.  We have reconstructed the spectra in each 20\,ks time
slice from a zero-point spectrum plus the first $p$ principal
components, $p=0,1,2...$, and measure the $\chi^2$ of the
goodness-of-fit of the reconstructed to the actual spectra.  The mean value of
$\chi^2$ averaged over all time slices is given in column four of
Table\,\ref{mkn766_2-10table}.  The mean spectrum on its own results in an
extremely large value of $\chi^2$; allowing the 1st principal
component reduces the $\chi^2$ such that the residual fluctuations
have an rms less than 10 percent larger than the photon shot noise; 
two variable components plus the zero-point spectrum render the PCA
model an acceptable fit within each timeslice.

However, we also consider the case where no constant zero-point spectrum
is allowed, with the PCA being carried out centred on the flux origin 
rather than the data mean.  As expected from the above, a single eigenvector
with no zero-point spectrum yields a rather poor fit to the data
($\chi^2=371$ for 105 degrees of freedom, column 7 of Table\,\ref{mkn766_2-10table})
despite describing 99 percent of the variance about the origin.
However, two eigenvectors, with no zero-point, provide an acceptable fit, in fact
almost as good as the case of a zero-point spectrum with two eigenvectors.  The
second eigenvector accounts of 0.6 percent of the variance and hence is substantially
smaller in its variation that the first eigenvector.

We infer from this analysis that the $2-10$\,keV data may be described by either:
(A) 
two independently-varying spectral components, one of which varies substantially
less than the other; or 
(B) a constant zero-point spectrum with one dominant
varying component, plus a further varying component that could either be a true
``third component'' or could be an indication at a low level of some more complex
variation such as would arise from variable absorption.  We shall investigate this
possibility further in section\,\ref{sec:0p4}.

\subsection{Limits on variation of the zero-point component}
It is not possible from such analysis to definitively state ``the zero-point component
is constant'' - a more meaningful statement is to place limits on the amount of
variation that could arise from its variation.  To make this measurement, we consider
model A from above, in which two variable components are allowed, and measure the
fractional variation required from the second component.  We choose the energy range
$2-10$\,keV to minimise the effect of any variable absorption. In principle, because of the
ambiguity introduced by allowing true physical components to be non-orthogonal,
it is possible to find a solution in which both components have large variations.
Accordingly, we choose to find the linear combination of eigenvectors that minimises
the required variation in the second component.  Effectively, this is measuring the
scatter about a best-fit relationship between the components plotted in 
Fig.\,\ref{mkn766_2-10keV_comp1_comp2}.  The r.m.s. fractional variation is found to
be 13\,percent, with a maximum of 37\,percent, so we can say that the zero-point
component is consistent with being constant to about this level.

\subsection{Results: 0.4-10\,keV}\label{sec:0p4}

We now extend this analysis to lower energies where we might expect any effects of
variable absorption to be more noticeable.  Repeating the above analysis assuming
the existence of a zero-point component, indeed we find that although a small number of
components give a broad description of the variations, a total of four variable components
plus the zero-point spectrum are required to reduce the mean $\chi^2$ in each timeslice to
the level expected from random noise (Table\,\ref{mkn766_0p4-10table}).
However, if we take this PCA but measure the contribution to $\chi^2$ only in the energy
range $2-10$\,keV we find two components are adequate, as found in the previous section.
This is partly because the shot noise errors at lower energy are substantially smaller than
at high energy, and hence small fractional departures from a good fit in the 
low energy channels have a relatively larger effect on $\chi^2$, but inspection of ratios
of the PCA reconstruction and the data show that the departures at lower energy are actually 
larger than at high energy.  The natural interpretation of the need for a large number of
additive components at low energies is that in fact there is variable absorption,
which cannot be correctly modelled by PCA.  In Paper\,II we measure 
the effects of variable absorption by fitting directly to timesliced data, a process which
confirms this phenomenon.

\begin{table}
{\small
\begin{tabular}{lr@{.}llrr}
& \multicolumn{4}{c}{0.4-10 keV} & 2-10 keV \\
\# & \multicolumn{2}{c}{$\lambda$} &
$\sigma_f^2$ & $\chi^2$/dof & $\chi^2$/dof \\
& \multicolumn{2}{c}{$/10^{-5}$} &  & \\
0 &  &   & 0.     & 41250/148 & 4942/106 \\
1 & 35&9 & 0.966  & 745/147 & 121/105 \\
2 & 0&35 & 0.0095 & 349/146   & 104/104 \\
3 & 0&13 & 0.0035 & 165/145   & 90/103 \\
4 & 0&088 & 0.0024 & 148/144  & 81/102 \\
5 & 0&080 & 0.0022 & 133/143  & 75/101 \\
\end{tabular}
}
\caption{
The amount of spectral variation explained by the first five 
principal components for Mrk\,766 in the energy range $0.4-10$\,keV.
The left-hand set of values provide the eigenvalues $\lambda$, fractional
variance $\sigma_f^2$ and $\chi^2$ values
for the analysis over this band, the right-hand columns give the contribution
to $\chi^2$ in just the $2-10$\,keV band from this analysis.
}
\label{mkn766_0p4-10table}
\end{table}

\subsection{Principal component spectra}\label{sec:PCspectra}
\begin{figure}
\resizebox{88mm}{!}{
\rotatebox{270}{
\includegraphics{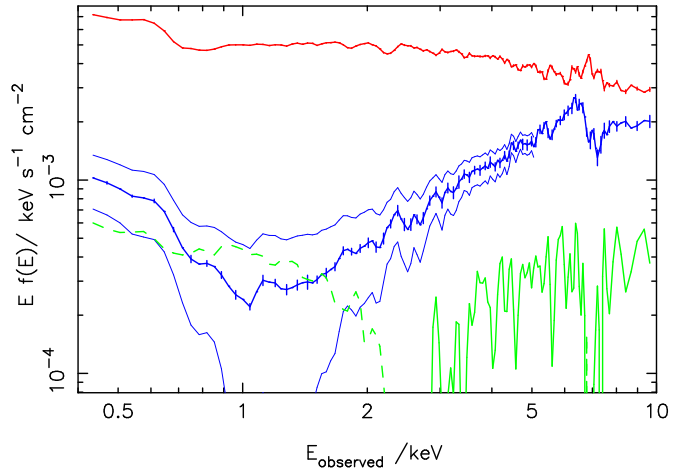}}}
\caption{
Principal component spectra of Mrk\,766, $0.4-10$\,keV.
The 1st principal component is the upper spectrum, scaled to have an amplitude equal to its maximum
above the mean spectrum.  
The possible range of the zero-point spectrum is shown by the lower set of spectra, 
with the upper and lower bounds discussed in the text
shown as light-weight lines and the mid-point between these shown in bold.
For clarity, these bounds are not plotted at energies above 5\,keV
where the random uncertainty becomes larger than this systematic uncertainty.
The true zero-point spectrum lies between these bounds,
but note that only systematic subtraction or addition of eigenvector one is allowed,
these are not random noise bounds.  The statistical uncertainties arising from
shot noise are determined as described in section\,\ref{sec:errors} and are shown
by the error bars on each spectrum (the errors on eigenvector one are very small).
The second principal component is also shown.
Errors on this component are large but are omitted for clarity.  
At low energies the component is negative:
the dashed portion of the curve shows the absolute value of the component.
}
\label{mkn766pcaspec}
\end{figure}

We now turn to inspection of the spectral components that arise from the PCA.  
Fig.\,\ref{mkn766pcaspec} shows the zero-point spectrum and first eigenvector
for case B, analysed over the $0.4-10$\,keV band.  As described above, the spectra
displayed in this section have been smoothed with a tophat function of FWHM 
approximately equal to the energy-dependent instrumental resolution, although all
analysis is done on spectra binned only into bins of width HWHM (section\,\ref{sec:generation}).
Eigenvector one is shown as the upper curve. Its amplitude varies of course, the
figure shows the maximum amplitude attained by this component in the PCA.  
The possible lower and upper bounds on the zero-point spectrum, discussed in 
section\,\ref{pca_introduction}, 
are also indicated by the pair of light-weight lines. The true zero-point spectrum must lie
between these extremes, and the 
main zero-point spectrum shown is simply the mean of the two extremes.
Note that only systematic subtraction or addition of eigenvectors is allowed.

We do not show the two eigenvectors required to describe the data for case A.
As these components are forced by the PCA to be orthogonal they will not necessarily
correspond to actual physical spectral components.  However, candidates for such
spectral components may be created from any linear combination of the two eigenvectors,
and in fact it is possible to construct two components whose 
spectral shape is indistinguishable from the two components already shown in 
Fig.\,\ref{mkn766pcaspec}.  In this description of the data, in effect,
the zero-point spectrum actually shows some variation, 
but with variability amplitude much smaller than that of eigenvector one.

There is no guarantee that the principal component spectra produced by this mathematical
decomposition will have any simple relation to actual physical emission components,
yet in the spectra we can immediately see physically-recognisable features.  
On eigenvector one, we can see that this component is dominated by a near power-law
in this energy range, but with a broad component of excess flux
covering the range $6-7$\,keV.  Its presence on eigenvector one implies that the line
emission varies with the continuum (i.e. its equivalent width is constant).
\citet{miller06} have identified a broad component of Fe\,K$\alpha$ emission from ionised Fe
in the band $6.5-7$\,keV that varies with the continuum on timescales $\ga 10$\,ks.
The PCA detection of this line on eigenvector one is in good agreement with the earlier
result.

The zero-point spectrum also shows clear, physically recognisable,
features.  First, the overall continuum shape is extremely hard, the
hard spectrum continuing up to the highest energies measured, but we
can also see a strong continuum break at $\sim 7$\,keV and a weak
component of emission at $\sim 6.4$\,keV: we identify these as the
Fe\,bound-free edge and Fe\,K$\alpha$ emission from low- or
medium-ionisation states of Fe. As well as the clearly-detected Fe
features, there are also indications of two discrete absorption lines
at energies both higher and lower than the nominal $7.14$\,keV break
rest energy.  All these features are discussed further in
sections\,\ref{modelling} \& \ref{discussion}.

We also see that the ``soft excess'' below 1\,keV that is
characteristic of many low-luminosity AGN is in fact present on both
the variable and constant components, implying that what we see is a
composite of different features.  Detailed modelling of the soft
excess will not be undertaken in this paper as it requires full
modelling of the variable absorption.

Finally we note the effect of higher-order eigenvectors.  If we
suppose that there are additive approximations to a true
(multiplicative) process of variable absorption, we can gauge the
extent to which variable absorption might corrupt the shape of our
zero-point spectrum and eigenvector one spectrum, by comparing the
amplitudes.  Their amplitude is such that they should have relatively
little effect on physical modelling of eigenvector one.  However, the
effect on the zero-point spectrum {\em could} be significant,
especially at energies $< 2 $\,keV.  We also note the presence of a
feature in eigenvector two at energies around 7\,keV: this might imply
that the possible absorption lines noted above are both variable and
tied to the continuum absorption variations that the rest of the
eigenvector is describing.

\section{Physical interpretation of the principal components}\label{modelling}

\subsection{Comparison with data}\label{sec:datacomparison}
\begin{figure*}
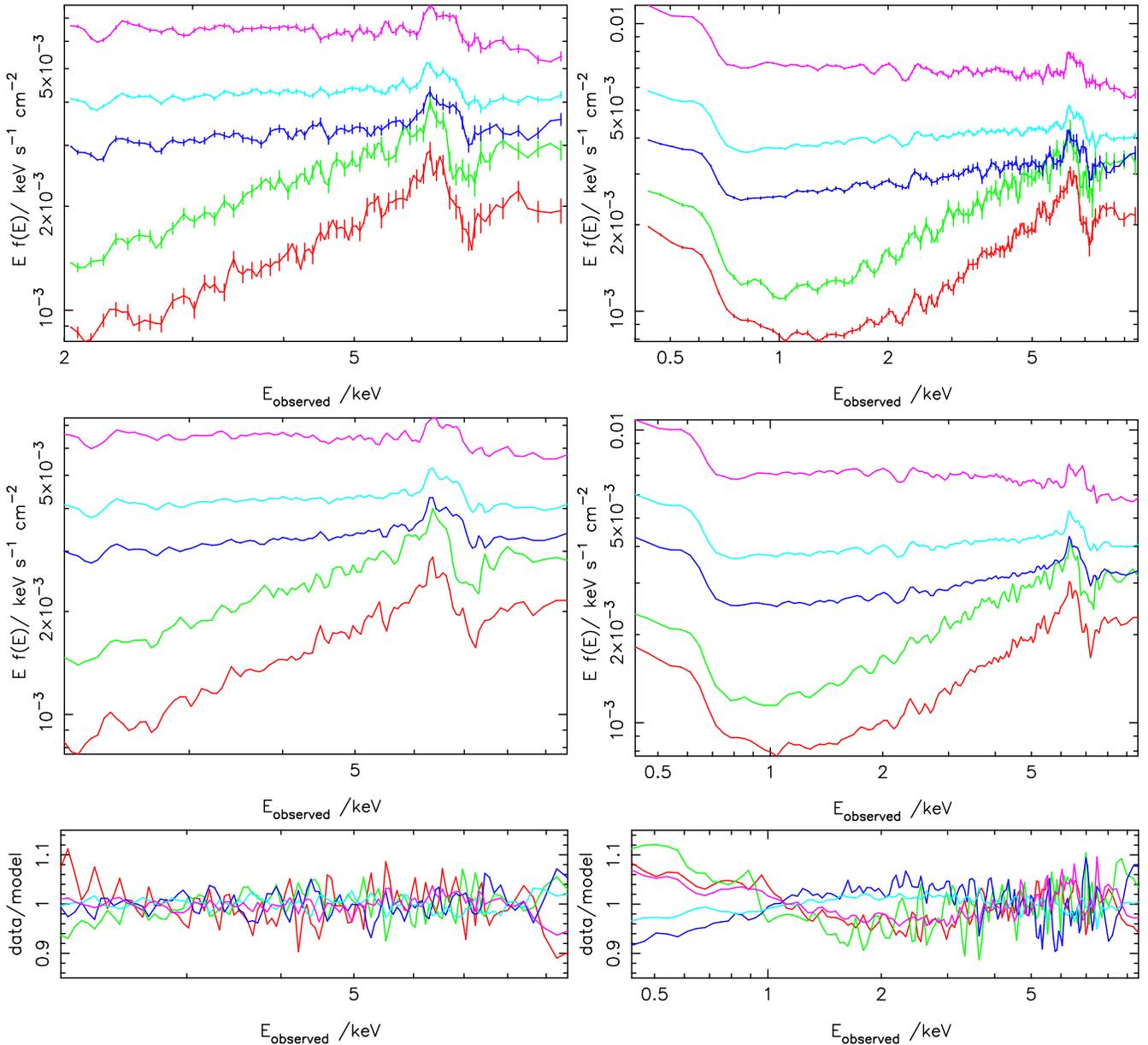

\begin{minipage}{88mm}{
\resizebox{88mm}{!}{
\rotatebox{270}{
\includegraphics{6548fig5a.ps}}}
}
\end{minipage}
\begin{minipage}{88mm}{
\resizebox{88mm}{!}{
\rotatebox{270}{
\includegraphics{6548fig5b.ps}}}
}
\end{minipage}

\begin{minipage}{88mm}{
\resizebox{88mm}{!}{
\rotatebox{270}{
\includegraphics{6548fig5c.ps}}}
}
\end{minipage}
\begin{minipage}{88mm}{
\resizebox{88mm}{!}{
\rotatebox{270}{
\includegraphics{6548fig5d.ps}}}
}
\end{minipage}

\begin{minipage}{88mm}{
\resizebox{88mm}{!}{
\rotatebox{270}{
\includegraphics{6548fig5e.ps}}}
}
\end{minipage}
\begin{minipage}{88mm}{
\resizebox{88mm}{!}{
\rotatebox{270}{
\includegraphics{6548fig5f.ps}}}
}
\end{minipage}

\caption{
Data used for the PCA, averaged into five flux states 
compared with the PCA reconstruction. {Left side:} $2-10$\,keV
analysis, flux states defined in the range $2-5$\,keV, 
{\em right side:} $0.4-10$\,keV analysis, flux states defined in the 
range $1-2$\,keV. 
{Top:} data averaged into the five
flux states;
{\em centre:} PCA reconstruction, using the zero-point spectrum plus the
first two eigenvectors;
{\em bottom:} ratio of data and PCA model (note the expanded y-axis scale).
}
\label{mkn766pcadata}
\end{figure*}

The above statistical analysis demonstrates that the data may be described by
a small number of variable components, but there is no guarantee that these
mathematical decompositions correspond to physical components.  
In the following sections we shall investigate a range of physical models that
may correspond to the principal components, but first we should investigate whether
the principal components actually yield a good description of the data.
Fig.\,\ref{mkn766pcadata} compares the PCA reconstruction with the actual data.
To reduce the effect of photon shot noise in this comparison, the data have been
averaged into five flux states, defined in equal logarithmic intervals of flux.
The left-hand figures show the $2-10$\,keV analysis of section\,\ref{sec:2-10}:
in this band the spectral variability is greatest at the lowest energies and the
flux integrated in the $2-5$\,keV band was used to define each flux state.
The right-hand figures show the $0.4-10$\,keV analysis of section\,\ref{sec:0p4};
here the variability is greatest in the $1-2$\,keV band and this was used to 
define the flux states.  In both cases the PCA reconstruction was made from 
the zero-point spectrum plus the first two eigenvectors.

It can be seen that the PCA models provide good agreement with the variations 
present in the data.  In the $2-10$\,keV range the fit is almost everywhere better than 
about 5 percent, apart from the extreme high and low energy limits.  
In the $0.4-10$\,keV range there is evidence for systematic
departures at the level of 10\,percent, especially at low energy, as expected from the
previous statistical analysis.
Overall, we conclude that the dominant spectral variability
is indeed well described by the simple additive model, although a detailed
description of the data will need to allow variable absorption.  The zero-point
spectrum and eigenvector one are sufficiently well-defined that we shall now
proceed to see if they can be fit by physical models of X-ray emitting regions in AGN.

\subsection{Model-fitting to Eigenvector one}
\label{sec:eigenvector}
We now fit physical models to the principal component spectra using the 
fitting software xspec
\citep{arnaud}. Because the PCA is a decomposition of the data, the component
spectra that are produced are convolved with the instrumental response function.
Hence these components may be treated as data for the purposes of fitting.
The actual instrument response of {\em XMM-Newton} varies with time, and the
PCA is based on observations obtained over the period 2000-2005.  
We use the instrument response functions from 2001, when the source was in its
high state.  Uncertainty in the true response
function causes some systematic uncertainty in the spectral fitting, which may be
largest for steep spectrum sources such as Mrk\,766.  Comparison of 
contemporary MOS and pn
observations of Mrk\,766 indicates this uncertainty could be as large as 5\,percent.  
There are
also systematic uncertainties arising in the generation of the PCA component
spectra: physical spectra may comprise any linear combination of the eigenvectors,
and when fitting over the full energy range $0.4-10$\,keV we have seen that a large
number of components are required to adequately describe the data.  Overall then, 
we should not expect a perfect fit between physical models and the component spectra.
In fitting to eigenvector one we allow a 3\,percent systematic error to be added in
quadrature with the random error when calculating $\chi^2$, although even this is
likely to be an underestimate of the true systematic error in some
regions of the spectrum.  One such region is the energy range $2.0-2.7$\,keV, which
is just above the energy of a sharp drop in the instrumental effective area.
Convolution of this sharp edge with the energy response function results in residual
ripples propagating into the principal components, visible in 
Figs\,4-5.  We exclude this spectral range when fitting eigenvector one.

\begin{figure}
\resizebox{88mm}{!}{
\rotatebox{270}{
\includegraphics{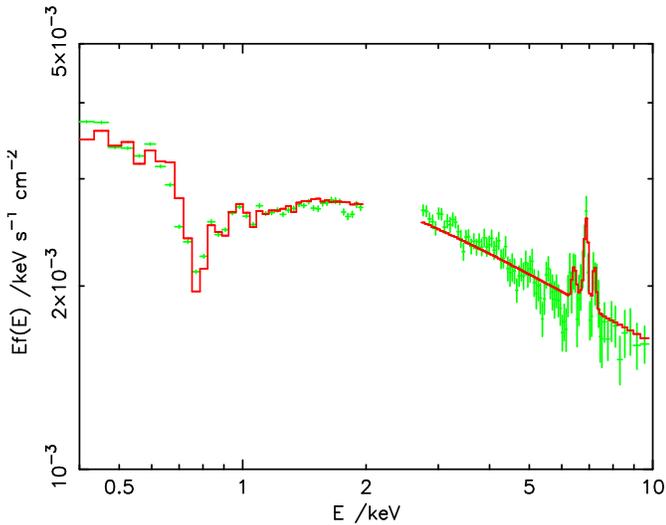}}}
\caption{
The fit to eigenvector one, with a model of an absorbed
power-law and some simple line components (see text).
The solid curve shows the model, points with errors the ``unfolded'' spectrum.
The spectrum in this plot is normalised to the value given by the eigenvalue,
i.e. to the eigenvector's rms amplitude.
}
\label{eigenvectorone}
\end{figure}

The model we fit to eigenvector one 
comprises two components.  The overall spectrum is assumed to
be a power-law affected by warm absorption: an absorption grid was created from
xstar \citep{kallman04} and its column and ionisation parameter were allowed to be
free parameters.  In the absorption models generated, the input source was assumed to be
a power-law of photon index 2.4.  The best-fit absorber parameter values were
$N_H = 2.7 \times 10^{21}$\,cm$^{-2}$
and $\log\xi=1$ (hereafter $\xi$ will be in units of \xiunit).
The ionised line reflection component was provided by a grid
of ``reflion'' models \citep{rossfabian}, with power-law index of the illuminating
radiation tied to that of the main power-law component, and with the reflection
ionisation as a free parameter.  
The reflection spectrum was convolved with a \citet{laor91} model using the xspec
routine ``kdblur''.
Additional warm xstar absorption was also
allowed on the reflection component, with best-fit values $N_H = 3.3 \times 10^{22}$\,cm$^{-2}$
and $\log\xi=1$.
This simple fit provides a good description of the overall spectral
shape, with a value of $\chi^2 = 142$ for 124 degrees of freedom, assuming a
3\,percent systematic error.
The model agrees with the component spectrum everywhere better than 10\,percent.
However, the fit is poor is around the 
broad Fe line,  and the line shape is
not well reproduced. In fact the data are better fit 
by a simpler model comprising only a warm-absorbed power-law plus a
line at 6.7\,keV blurred by the same Laor model as above: i.e. without including
the reflected continuum and Compton broadening that we expect to be associated with 
ionised reflection.  The best-fit value of power-law index in steep,
$\Gamma=2.38 \pm .01$ (statistical error) $\pm .05 $ (estimated systematic error obtained
from fitting to higher energies alone).
Additional narrow components of ``emission'' at rest energies
of 6.97 and 7.3\,keV, {\em not}
blurred by an accretion disc, improve the fit further: we discuss the interpretation of these
components below.
The fit is shown in Fig.\,\ref{eigenvectorone}: $\chi^2$ has now 
improved to 116 with 124 degrees of freedom.
The range of accretion disc radii responsible
for the ionised line is not well-determined: 
the best-fit value for the inner radius of the emission assuming a disk emissivity
power-law index of 3 and an inclination angle of $30^{\circ}$
is $r_{\rm in} \simeq 100$r$_g$, but no useful upper or lower limit
on this value is defined by the fit to the line profile, largely as a result of allowing
the additional narrow line components in the fit.

The two additional ``emission'' components occur at the energies of likely
absorption lines identifiable in the zero-point spectrum, at rest energies
6.97 and 7.3\,keV, which are discussed below.  It seems likely that the 
components visible on eigenvector one are in fact indicative of those absorption
lines disappearing as the source brightens, rather than being a genuine 
emission component.
Overall, we conclude that the principal varying component, eigenvector one,
is likely to comprise a variable power-law component and an associated broad
reflected ionised Fe line
to be discussed further in section\,\ref{discussion}, although
the actual lineshape may be contaminated by the effects of varying absorption.
The worse fit of the reflion model implies that the
reflected {\em continuum} expected to be associated with the ionised
line has however not yet been clearly detected through its spectral shape.

\subsection{The zero-point spectrum}
An extremely hard spectrum and strong Fe\,K edge as seen in the zero-point
spectrum might arise from either  
a component of ``reflected'' emission from neutral or moderate-ionisation
optically-thick gas \citep{georgefabian, magdziarz95, rossfabian}, or from
ionised absorption of the continuum source.
If reflection arises from the inner regions of an accretion disc it may 
also be relativistically blurred. In this section we attempt to fit a number of 
possible physical models to the zero-point spectrum. From the analysis
in section\,4 it seems likely that this component is significantly affected by 
variable absorption below $\sim 1-2$\,keV, and hence in this paper we concentrate on fitting
to the energy range $1-10$\,keV.  In fact, the most important diagnostics for
interpreting the physical origin of this emission are the Fe\,K emission line
and absorption edge.  Two features are particularly noticeable. First, the absorption
edge, although having a large decrement, is soft and appears blurred or absorbed to
energies lower than the canonical (rest) energy of 7.14\,keV expected from neutral
material \citep{kallman04}.  Second, the equivalent width of neutral 6.4\,keV
Fe\,K$\alpha$ is much weaker than expected if this component is identified with 
pure reflection from low ionisation material: the observed
equivalent width is $\sim 50$\,eV compared with the expected value $\sim 1$\,keV
\citep{georgefabian}.

There are a number of possible explanations for the low equivalent width and apparently
soft edge,
and we discuss some of these in turn by fitting an appropriate model to the
zero-point spectrum.  We noted in section\,\ref{pca_introduction} that this component
is not uniquely defined: the actual zero-point spectrum could have any amount 
of eigenvector one added to it in order to produce the set of possible zero-point 
curves shown in Fig.\,\ref{mkn766pcaspec}.  We deal with this 
in the following manner. We choose to fit to the zero-point spectrum deduced from the
$0.4-9.8$\,keV PCA that is the maximum of the allowed extremes
shown in Fig.\,\ref{mkn766pcaspec}.
Then, when fitting physical models to this spectrum, we allow also a variable amount
of eigenvector one to be included in the fit.  This process thus exactly mimics the
systematic uncertainty in the definition of the zero-point spectrum.  The only
complication is that eigenvector one is defined as data, convolved with the instrumental
response, so in order to correctly include this in the fitting we allow an
arbitrary amplitude of the model fit to eigenvector one, 
rather than eigenvector one itself.  We find that no component of systematic error
is required in fitting the zero-point spectrum in the $1-10$\,keV range.  
We also do not exclude the range
$2.0-2.7$\,keV as we did for eigenvector one: 
the larger random uncertainties on the zero-point spectrum mean that
the systematic errors on the component spectrum in this energy range do not significantly
affect the values of $\chi^2$ obtained.

\subsubsection{Absorption lines in the zero-point spectrum}
\label{abslines}
Inspection of Fig.\,\ref{mkn766pcadata} indicates the presence of two 
absorption features, possibly lines, at observed energies about 6.9 and 7.2\,keV,
most prominently in the lowest flux states.  These features also appear in the 
PCA zero-point spectrum, as shown in Fig.\,\ref{mkn766pcaspec}.  Whether these
features are actually absorption lines does depend on the model adopted for
the K-edge and continuum, so their statistical significance is assessed in
each of the specific models discussed below.  They appear primarily on the
zero-point component spectrum because their absorbed flux is approximately constant:
absorption lines of constant equivalent width would appear on eigenvector one.
The most natural interpretation of their
appearance on the zero-point component is that these are indicative of an
absorbing region that is physically localised to the region represented by the
zero-point component.  Having said that, the positive features that
appear at these energies in
eigenvector one, and also in eigenvector two, may indicate a more complex 
variability behaviour for these absorption lines.

If the lines are real, their identification is
ambiguous.  The feature observed at 6.9\,keV could be 6.97\,keV\,Fe\,XXVI\,Ly$\alpha$
in the rest-frame of the source, or it could be Fe\,K$\alpha$ absorption with
some outflow velocity ($\sim 13,000$\,km\,s$^{-1}$ for 6.7\,keV\,K$\alpha$ or
$\sim 22,500$\,km\,s$^{-1}$ for 6.5\,keV\,K$\alpha$).
The feature at 7.2\,keV could be 
7.3\,keV\,Fe\,XIX\,K$\beta$ in the source rest-frame, although in this case
we would expect to see a broad complex of Fe absorption in this region of
the spectrum \citep{kallman04}.  Alternatively, it could be blueshifted
absorption: 6.97\,keV\,Fe\,XXVI\,Ly$\alpha$ at $\sim 13,000$\,km\,s$^{-1}$
or Fe\,K$\alpha$ absorption 
($\sim 26,500$\,km\,s$^{-1}$ for 6.7\,keV\,K$\alpha$ or
$\sim 37,000$\,km\,s$^{-1}$ for 6.5\,keV\,K$\alpha$).  Any of these cases are
indicative of a high column density of ionised ($\log\xi > 2$) gas.  
The $13,000$\,km\,s$^{-1}$ solution is interesting as it might imply a common origin
for both lines.  Further discussion of the identification of these features 
and their variability is postponed to Paper\,II: in the following sections
we allow for their possible presence by including Gaussian absorption lines
of variable amplitude at these two energies.

\subsubsection{Model components for the zero-point spectrum}
In the following we present a number of alternative model fits to the zero-point
spectrum.  Each model has the same general structure, although the detailed
components vary.  

First, we assume that the low-column, low-ionisation absorption
that was identified in the fit to eigenvector one is also an absorption layer
in front of the zero-point spectrum.  The parameters of this layer are frozen
at the values found above.
Then, the zero-point spectrum is allowed to comprise three components:
(i) some variable amount of eigenvector one, as discussed above;
(ii) a component of ionised reflection or absorption (depending on the model below)
whose incident illumination is a power-law of the same slope as eigenvector
one, the slope being a fixed parameter; and 
(iii) a component of neutral reflection with an unresolved 
6.4\,keV\,Fe emission line of equivalent width 900\,eV with respect to the reflected
continuum, a typical value
expected for a steep-spectrum reflector \citep{georgefabian}.  
Component (iii) is modelled within xspec using the ``pexrav'' function
\citep{magdziarz95}.
We also allow an additional component of ionised absorption to be in front of 
component (ii), the ionised reflection/absorption:  this might arise if the
reflecting surface has its own atmosphere, for example.  

Although fairly complex, this is probably a minimal set of components for a 
physical model of the emitting regions, and
the quality of the data and principal component spectra that we have extracted
do demand fitted models of this complexity.  Variants on the above general model
are easily envisaged (in particular, the choice of which absorbing layers cover
which emission regions) but in the fitting below we keep the
same general structure for all models in order to allow fair comparison between them.

Finally, we note again the important caveat that we are here not fitting directly to
data but to the PCA zero-point spectrum.  Depending on the interpretation of
eigenvector two, in terms of models A or B of section\,\ref{sec:2-10}, the zero-point
spectrum below 2\,keV may either be significantly affected by variable absorption (model B)
or not (model A).  Thus too much reliance should not be placed on the parameter
values that are found when fitting to the zero-point spectrum: the important point
is to find out whether physical models can be found which may in principle explain the
existence of the zero-point spectral component, and if so what range of models may
suffice.

\subsubsection{Relativistically-blurred disc reflection}\label{sec:blur}
A similar component to Mrk\,766's zero-point spectrum arises also in MCG--6-30-15
\citep{vaughanfabian04}, and in that case the spectral shape is thought to
arise from relativistic blurring of reflection from an accretion disc.
We investigate whether such an explanation is viable in Mrk\,766, first by
fitting to the zero-point spectrum in this section, and then discussing the required
geometry of such a model in section\,\ref{discussion}.

To create a model of blurred reflection, we take the ionised reflection  
``reflion'' model of \citet{rossfabian} and blur it with
a \citet{laor91} convolution model, provided in xspec by the ``kdblur''
function: a layer of ionised absorption is also allowed as described above.
This then is used as component (ii) in fitting, together with 
eigenvector one and the neutral reflection components. 
An important feature in the ionised reflection
component spectrum is the depth of the Fe\,K edge,
which is determined by both the abundance of Fe and the geometry of the reflection.
The latter is not a variable in the reflection model, but we can allow the Fe
abundance to vary. For consistency we scale the pexrav model and its associated
Fe\,K$\alpha$ line to the same abundance value.
The best-fit value was found to be 0.35 times solar:
this model provides a good fit to the spectrum over the range $1-10$\,keV
(Fig.\,\ref{fig:zeropoint}) with $\chi^2 =96$ for 122 degrees of freedom, with
10 free parameters.  
We note here that this fit is much better than expected given
the assumed size of error bars.  We believe this is because of neglect of covariance
between spectrum points, so that in effect the true number of degrees of freedom
is smaller than assumed.  Nonetheless, the fit is clearly good.
Increasing the Fe abundance to 0.5 increases $\chi^2$ to
a value 103, forcing Fe abundance to have the solar value results in $\chi^2=162$.
The maximum departure between the $1-10$\,keV spectrum and the best-fitting
model is $\sim 15$\,percent.
The best-fit reflion ionisation parameter is $\xi\simeq 700$\xiunit with an
absorbing layer of column $N_H = 3 \times 10^{22}$\,cm$^{-2}$ and $\log\xi\simeq -1$.
The relativistic blurring has a major effect in this model, the 
inner radius of the emission having a best-fit value equal to
the minimum allowed value in the model of r$_{\rm in}=1.235$r$_g$.
The unblurred neutral reflection
is an important component in allowing the blurred model to fit the data: without
it $\chi^2$ increases to a value 480, even allowing the remaining parameters to vary.
The flux in the $1-10$\,keV band provided by the blurred component is
$3.3 \times 10^{-12}$\,erg\,cm$^{-2}$\,s$^{-1}$ and by the neutral component
is $0.5 \times 10^{-12}$\,erg\,cm$^{-2}$\,s$^{-1}$.
The two absorption lines
discussed in section\,\ref{abslines} also appear to be significant in this model:
$\Delta\chi^2$ is found to be 40, 44 or 74 in the absence of the 6.9\,keV, 7.2\,keV
or both absorption lines, respectively, so in the fitted model a substantial part of
the softening of the Fe\,K edge is caused by the presence of absorption.

Is relativistic blurring required?  Increasing r$_{\rm in}$
even to a value 6r$_g$ increases $\chi^2$ by 23, so it seems that, in the reflion
model fitted here, relativistic blurring is strongly favoured.
It is possible to obtain a reasonable (but not good)
fit of such an optically-thick ionised-reflection 
model to the zero-point spectrum without requiring relativistic
blurring, provided that the iron abundance is allowed to be low: a model comprising 
neutral reflection plus absorbed ``reflion'' reflection with 
ionisation parameter $\ga 1000$\xiunit\, both with Fe abundance
0.13 times solar and covered by an ionised absorber can fit the component spectrum
with $\chi^2=160$ with 9 free parameters and 123 degrees of freedom in the
$1-10$\,keV range.  In fact, we expect the equivalent width of the emission lines
to be geometry dependent in optically-thick reflection: 
the apparently low Fe abundance may instead be a signature
of a different geometry from that assumed in the \citet{rossfabian} model.
However, in the following sections we consider whether more extreme variations
from the assumptions of the reflion model may provide alternative explanations
of the zero-point component without requiring relativistic blurring.

\subsubsection{Non-relativistically blurred reflection models}
\label{sec:lowEWreflection}
In this section
we investigate alternative reflection scenarios which might be able 
to explain the spectral shape without requiring relativistic blurring.  In this
case the reflecting material could be placed at larger distances from the central
illuminating source, allowing the lack of variability to be explained as arising
from the light travel-time delay.  The critical question becomes, ``can the strong Fe 
K-edge but weak Fe\,K$\alpha$ emission line arise naturally in reflection models?''.
The low equivalent width of 6.4\,keV\,Fe\,K$\alpha$ makes it infeasible that the 
zero-point spectrum may be fit by a standard low-ionisation reflection model.
It seems likely therefore that, regardless of whether it is relativistically
blurred or not, if there is a substantial component of reflection then it is 
from Fe-ionised material.

\begin{figure}
\resizebox{88mm}{!}{
\rotatebox{270}{
\includegraphics{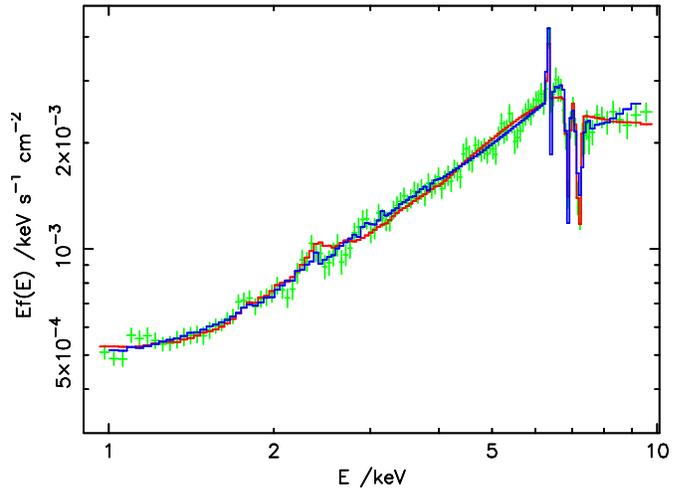}}}
\caption{
Fits to the zero-point spectrum showing the relativistically-blurred model
(red curve, lowest of the two models at 10\,keV) and the 
non-blurred ionised-opacity model (blue curve)
with the unfolded data points, with errors.
}
\label{fig:zeropoint}
\end{figure}

There are three ways in which ionised reflection might be able to generate a spectrum
close to that observed.
The first is to suppose that the reflector is a high column, close to being optically-thick,
of very highly ionised gas that reflects radiation from the central X-ray source through
Compton scattering off free electrons.  In order to obtain a reflected intensity close
to that observed the scattering medium would need to cover a substantial fraction of the 
source: the evidence in favour of such a model is discussed in section\,\ref{discussion}.
The observed spectral shape would be produced by a high column
of some separate intervening absorption of intermediate ionisation.  The existence of
an outer layer of lower ionisation material might be a natural
expectation for a region of gas photoionised by a central source. 
We therefore model component (ii) in our generic scheme 
as being a component of absorbed scattered radiation
whose shape is parameterised as an absorbed component of eigenvector one 
with absorbing column and ionisation
parameter being free parameters.  We compared absorber models
with two values of Fe abundance, solar and 0.67 times solar: the latter provides a
significantly better fit (note that we also had to allow low Fe abundance in the
blurred reflection model).  In principle we could have allowed Fe abundance to be a free
parameter in the fitting, but it was somewhat prohibitive to generate a wide range of xstar
absorption models: in practice we find that the choice of 0.67 already provides an excellent fit
to the data, so investigation of a wider range of abundance was not justified. 
The best fit values for the absorber were found to be
$\log\xi=3.0$, $N_H=4\times 10^{23}$\,cm$^{-2}$: 
For consistency with the other models discussed here, this
component is also covered by a low-ionisation absorbing layer which has best-fit values
$N_H = 5 \times 10 ^{22}$\,cm$^{-2}$, $\log\xi \simeq 0$.
The fit of this model, shown in Fig.\,\ref{fig:zeropoint},
is actually better than in the case of blurred ionised reflection: 
$\chi^2 = 83$ with 9 free parameters and
123 degrees of freedom (although we should decrease the number of free parameters by one to account
for the freedom of choosing a non-solar abundance).
Again, the pair
of absorption lines are required in this model, they play a significant role in defining
the shape of the edge around 7\,keV. The best-fit model does predict some amount of
6.5\,keV\,Fe\,K$\alpha$ seen in absorption: if we require
a value for the ionisation parameter $\log\xi < 2.5$ in order to avoid any $6.5$\,keV
absorption, we find $\chi^2$ to be slightly
higher, with a value $94$ for 124 degrees of freedom, 
but still a very good fit to the data.

A second way of obtaining the observed spectrum could be from a reflector
of intermediate ionisation, Fe ionised in the range Fe\,XVIII-XXIV,
such that the K$\alpha$ line can be resonantly scattered and destroyed by the Auger effect
(\citealt{ross,matt96,zycki94}, see also \citealt{liedahl}).
Resonant Auger destruction arises in the ``reflion'' models of 
\citet{rossfabian} and has already been suggested for 
a model incorporating a reflection component in Mrk\,766 by \citet{matt00}.
However, we have already seen in section\,\ref{sec:blur} that the optically-thick
reflion models do not provide the best fit to the spectrum without
relativistic blurring. The reason is that in the reflion model 
with $\xi \sim 300$\xiunit the opacity of
the gas within the reflecting region falls sufficiently low that strong emission
lines from lower ionisation material, particularly 6.4\,keV\,Fe\,K$\alpha$, are produced.

As an example of a model that might avoid this problem, consider a region of reflecting gas
illuminated from within by the central X-ray source of the AGN, such as 
a wind or atmosphere above the accretion disc being illuminated by a central X-ray
source.  The distant observer would have both a direct view of the central source
and a view of reflected radiation from the wind or atmosphere.
If the density within the
scattering region falls off with radius, perhaps as fast as $r^{-2}$, the
ionisation state of the gas could be high throughout, diminished only by the 
opacity of the gas (as in the plane-parallel reflecting slab). However, the
column of gas may be only just Compton-thick, perhaps $N_H \la 10^{24}$\,cm$^{-2}$,
so that the gas opacity does not become too high: if the source covering fraction
is high, 
a scattered intensity comparable to that observed in Mrk\,766 could be
produced without having a high equivalent width in either 6.4\,keV or 6.5\,keV
Fe\,K$\alpha$ line emission.  

To correctly model the expected spectrum from such a scattering region would require
a radiative transfer code similar to that of \citet{rossfabian},
incorporating the effects of resonant scattering and Auger destruction,
constructed with a relevant geometry, far beyond the scope of this paper.  
However, if the range of ionisation parameter within the
gas is small we can approximate the resulting spectrum by considering the effect 
of absorption on an incident spectrum, and we might expect that again 
model component (ii) may have a spectrum 
whose shape is parameterised as an absorbed power-law.
This is approximately true in the reflion optically-thick slab model:
in this $\xi$ range, the effect of Compton energy redistribution 
on the spectrum shape is small, and the continuum shape in the 1-10\,keV
range is dominated by the gas opacity within the reflecting zone.  
In this case, we would 
require the range of ionisation to allow resonant Auger
destruction, i.e. $2.5 \la \log\xi \la 3.5$ for photon spectral index $\Gamma=2.4$.
One sophistication that would differentiate the ionised reflection component
from a simple absorbing
model is that we assume the high-ionisation zone is responsible for scattering or
reflecting the incident radiation: in this case both the expected 6.5\,keV\,Fe\,K$\alpha$
line and the continuum are assumed isotropically scattered, and hence no 
6.5\,keV absorption line should appear in its spectrum.  
This would not be the case if we were viewing a
background radiation source in transmission through the ionised gas, when we would
expect Fe\,K$\alpha$ to be seen in absorption.  If we artificially remove 
6.5\,keV\,Fe\,K$\alpha$ absorption by allowing a cancelling emission line in the model
fit, a slightly higher ionisation parameter is favoured, $\log\xi \simeq 3.1$, with
an improved $\chi^2=78$ with 122 degrees of freedom.  We emphasise that a 
proper evaluation of this suggestion does require construction of a reflection model
incorporating the full range of opacity expected, given the chosen geometry.

Finally, the observed
Fe\,K$\alpha$ emission may be further suppressed geometrically in models of reflection
from extended regions: because the Fe\,K$\alpha$
line opacity is much higher than the continuum opacity, the observed Fe\,K$\alpha$
photons are effectively seen from a relatively thin outer scattering layer: if the
region is arranged to have greater surface area oriented away from the observer,
substantially reduced observed line emission may result \citep{ferland92}.  

\subsection{Absorption-only models} 
Given the success of the absorption fits in the
preceding section, we consider whether in fact models comprising only absorption, and no
reflection, may explain the data.
In order to explain the time-variation in spectral shape, either
the opacity must vary, or the emitting source must be only partially covered
by an absorber whose covering fraction varies.  Crucially, any such model
must satisfy the requirement that the resulting spectral variation yields the
{\em appearance} 
of arising from additive components, in order to yield the observed
PCA results.
It thus seems highly unlikely that in fact the opacity could vary in just such a way
to mimic the effect of two primary additive components in the data, as displayed
in Fig.\,\ref{mkn766_2-10keV_comp1_comp2}, throughout a range of a factor 10
in flux at $\sim 2$\,keV. We do not consider such variable-opacity models further
here.

The partial covering model seems more viable however: in this case the observed spectrum
again comprises two additive components, and because the true components may be
formed from any linear addition of the components deduced in section\,\ref{sec:pca}
such a model must fit the data equally as well as the reflection models considered
above.  In essence, the time-dependent source spectrum would be a linear combination of
an absorbed component that dominates in the low state and an unabsorbed component
that dominates in the high state, with a smooth progression in the relative contribution
of each between these states.
The only distinguishing feature might be that in this case we might expect to
see Fe\,K$\alpha$ absorption from the ionised absorber that is required to fit the
spectral shape; no absorption in the rest-frame is detected, although
one of the lines at 6.9 or 7.2\,keV could be blueshifted Fe\,K$\alpha$.

\citet{pounds04} have also suggested 
a similar scenario for the Seyfert\,1 AGN 1H\,$0419-577$, and argue that its
spectral variability may be explained either as a component of extremely
blurred inner-disc emission or as absorption partially covering the 
continuum source.  

The physical constraints on such a model are interesting: in this
picture {\em the factor ten variations in source flux are caused by
the covering fraction variations}.  The source flux varies extremely
rapidly, with a break in the power spectrum at a frequency $\sim 5
\times 10 ^{-4}$\,Hz \citep{vaughan03,markowitz}.  This is the orbital
timescale at $\sim 10$r$_g$ for black hole mass $4 \times
10^6$\,M$_{\odot}$.  However, the analysis presented here has only
considered spectral variations on timescales as short as 20\,ks, so we
might suppose that some other process produces the extremely rapid
variability, but that the partial covering varies on periods $\sim
40$\,ks.  This timescale corresponds to the orbital timescale at $\sim
70$\,r$_g$.  Either the absorption would arise from this scale, in
order to produce the observed time variability, or the emitter is both
extended and inhomogeneous on this scale, producing variability by
moving behind a patchy absorber.  

An intriguing example of an absorber model
has been suggested by \citet{done}, who propose that an outflowing disk
wind should produce an edge-like P-Cygni profile: they show that such a
model fits the spectral shape of 1H\,$0707-495$. Interestingly, the model
can produce both an edge at about the energy of the atomic transition
considered (\citealt{done} assumed 6.95\,keV) and also absorption
at higher energies.  It may be that the overall shape of the edge and 
7.2\,keV (in the observed frame) absorption
seen in Mrk\,766 may also be explained by such a model.  If the observed-frame
6.9\,keV absorption is real, however, this might need an additional 
non-outflowing absorbing layer.  Models corresponding to this proposal
are discussed by \citet{done} but 
are not yet generally available, so fitting of such a model is postponed
to future work.  Observationally, we might expect absorption and reflection
models to have different signatures at very hard energies near the ``Compton hump'',
and this might be testable with Suzaku observations, although \citet{done}
point out that high columns of absorbing material would in any case lead to
significant reflection also.

\section{Discussion}\label{discussion}
\subsection{The principal varying component}
It seems very likely that the primary variable component in Mrk\,766 has
the form of an absorbed power-law, with accompanying ionised reflection
that varies with the continuum.  This conclusion is in direct agreement
with the conclusions of \citet{miller06} who measured the correlation
between ionised line emission and continuum variations in the same dataset.
The index of the power-law is consistent with being unchanging over the four
year period, and the goodness-of-fit of a small number of PCA components
implies that substantial variation in spectral index (see e.g. \citealt{poutanen})
does not seem to be occurring in Mrk\,766.
In the analysis presented here, measurement of the ionised line profile appears to be 
contaminated by the effect of variable absorption, but the line width does
appear consistent with an origin at about 100\,r$_g$, consistent with
the analysis of the 2001 data by \citet{turner766} who found evidence for
periodic Doppler shift of ionised emission from a radius $r \sim 100$r$_g$
(assuming a disc inclination $\sim 30^{\circ}$).  Thus the three analyses 
yield a consistent location for the line emission.

\subsection{The zero-point spectrum}
The physical interpretation of the zero-point spectrum is however more
ambiguous, and no definite conclusion can yet be reached on its origin.
It may be that the blurred K-absorption edge and weak 6.4\,keV line are
a result of relativistic blurring with an inner radius for the emission
$<6$r$_g$.  However, this leaves a number of puzzles.  This component appears
constant to within 37\,percent, with an rms fractional variation of 13\,percent,
over the period 2000-2005, despite
variation in the illuminating continuum of a factor 10.  A similar dilemma
has been found for MCG--6-30-15, which has many similarities with Mrk\,766
\citep{miniutti03,vaughanfabian04}.  
To explain this behaviour whilst retaining the inner-disc
hypothesis, two possible explanations have been proposed.  

The first
invokes gravitational bending of light from an illuminating hotspot around
the black hole, in such a way that the continuum from the hotspot appears
to vary while the reflected emission from the disc remains approximately
constant \citep{miniutti03,miniutti04}. Given the result of \citet{miller06},
confirmed by the PCA, 
the reflecting accretion disc at $\sim 100$r$_g$ also needs to see the
same apparent continuum variation as the distant observer, because the
reflected emission varies closely with the continuum.  It may be difficult 
to create a geometry that would allow this, especially if the phenomenon of
a constant reflected component is a general feature seen in many AGN.

\citet{merloni06} propose an alternative explanation for the lack of variation of inner-disc
emission.  They predict
the emission expected from an unstable radiation-pressure-dominated
accretion disc, which has both density and heating inhomogeneities, and
conclude that variability of the reflected emission could be
suppressed in this model.  It is not yet clear however whether the variation
over long timescales 
would be as small as observed in Mrk\,766, nor whether the variability
could be reinstated at 100r$_g$.  The role of the accretion disc in
suppressing variation in reflection has also been discussed by
\citet{nayakshin}, who find that reflection variations may be suppressed
on timescales comparable to the disc dynamical timescale.  However, if 
the reflected emission originates at only a few gravitational radii, the
dynamical timescale is expected to be much shorter than sampled by our
analysis.  In this case we would still expect to see longer-term variations
in the reflected emission which are not observed, so this effect
seems unlikely to explain the apparent near-constancy of
the zero-point component on timescales $>20$\,ks.

An alternative interpretation of the lack of variability in this component 
is dilution of variations by light travel-time, with emission coming
from regions significantly more distant than the inner accretion disc.
However, we then need to explain the lack of Fe\,K$\alpha$ emission.  In the
model fitting, we considered two possible scenarios for a ``distant reflector''
hypothesis.  In the first, the reflector
was assumed to be a high column of highly-ionised gas covering a large fraction
of the source, with a separate absorbing layer being responsible for the
hard spectrum observed.  
In the second, the hard spectrum could arise
from the opacity of the reflecting material itself, rather than being an
absorbing layer that is separate from the reflecting medium.
In the second case Fe\,K$\alpha$ emission would need to be suppressed either
geometrically or by resonant Auger destruction.  

The existence of such gas may be expected from models of the central regions, and
\citet{kingpounds} have proposed that optically-thick winds should be expected
from high Eddington-ratio AGN.
Ionised gas of density $\sim 10^7$\,cm$^{-3}$ would
be Compton thick if extended over a region larger than $\sim 10^{17}$\,cm: such a
region would have a light-travel-time radius of about one month, sufficient to explain
the lack of variability in the reflected spectrum.
In the model of a disc wind associated with a $10^8$\,M$_{\odot}$ black hole
by \citet{proga04}, $\sim 80$\,percent of the source is covered
by a column of $N_H \sim 5 \times 10^{23}$\,cm$^{-2}$ gas with $\log\xi > 4$ out
to the largest radius simulated in their model of $\sim 5 \times 10^{16}$\,cm, and this
region would not need to be much larger in order to achieve a Compton scattering
depth about unity.
There is also observational evidence for such gas: the existence of significant columns
of ionised absorption has long been recognised (e.g. \citealt{kraemer, turnerea05} and 
references therein). In our X-ray observations
of Mrk\,766 the most plausible identifications of the twin absorption lines 
observed at $6.9$ and $7.2$\,keV require a high column ($\ga 10^{23}$\,cm$^{-2}$,
dependent on line identification and velocity dispersion of the gas) of high
ionisation (possibly as high as $\log\xi \sim 5$ depending on line identification)
gas.  Confirmation of this zone should be a high priority for future spectroscopic
observations.  High-column outflowing winds have also been proposed to explain 
observed absorption features in other AGN \citep{reeves03, pounds03b}.
There is also evidence from optical polarisation studies of AGN 
that a high column of ionised material is responsible for scattering above the
accretion disc at radii comparable to the broad-line region: \citet{smith} infer
a column $\sim 4 \times 10^{23}$\,cm$^{-2}$ between radii $1.5$ 
and $2.4 \times 10^{17}$\,cm, 
extrapolation to smaller radii would imply the existence of an absorbing layer that is
Compton-thick to radiation from the central source, albeit with unknown covering
fraction. We have therefore 
suggested that reflection from an extended atmosphere or wind may provide
a spectrum similar to that deduced from the PCA.
Higher resolution spectroscopy of the region around 7\,keV should again allow the
possible models to be distinguished.

\subsection{The absorption lines}
Given the limited resolution of {\em XMM-Newton} and the complexity of the spectra
around the Fe\,K-edge, it is difficult to be certain that the absorption features
at 6.9 and 7.2\,keV are discrete lines.  Both models considered here do require them
however.  The lines are interesting in themselves, as they may be indicative of
high-velocity, high-ionisation outflow, as mentioned above.  Both lines may be produced by gas with
$\log\xi \sim 4$ with outflow velocity $13,000$\,km\,s$^{-1}$: interestingly
gas can acquire such radial velocities in the disc wind model of \citet{proga04}
discussed above. High velocity H- and He-like Fe absorption also appears in the model
of \citet{sim05}, which aims to explain the apparent high outflow velocity in 
PG\,$1211+143$.  High ionisation outflows have also been detected in Galactic X-ray
sources (e.g. \citealt{kotani}, \citealt{jmiller}).
In Mrk\,766, higher resolution
data would help clarify the nature of these features, although it does appear that
they are only strong in the low state.  It is possible that these absorption
features are symptoms of a much more dramatic absorption signature, in which the
entire Fe\,K edge is produced by P-Cygni-like absorption from a disc wind, as suggested
by \citet{done}.  In this case the wind would need to have a variable covering fraction
in order to reproduce the observed spectral variability.  The variability timescales
suggest a wind origin again $\sim 100$r$_g$.

\section{Conclusions}

Our main conclusions may be summarised as follows.
\begin{itemize}
\setlength\itemsep{0mm}
\item Principal components analysis reveals that a relatively simple additive
model can explain the complex spectral variability of Mrk\,766: namely that there is a
variable component dominated by a soft power-law component, of constant power-law index, 
together with relatively non-variable component(s) constant to
$\sim 40$\,percent.

\item A variable ionised Fe emission line is detected on the principal
varying component, eigenvector one, that is probably high-ionisation
reflection from the inner accretion disc, consistent with the analysis
of \citet{miller06} of the same dataset.  The best-fit models indicate
an origin at $\sim 100$r$_g$ for the ionised Fe emission assuming a
disc inclination angle $\sim 30^{\circ}$.

\item The relatively unchanging zero-point component has a hard spectrum with a pronounced
Fe\,K edge but only weak Fe\,K$\alpha$ emission, whose spectrum seems to require
ionised reflection or absorption.  The Fe\,K edge appears soft and
is likely affected either by a combination of relativistic blurring and absorption, or by
absorption alone.

\item
If the zero-point component originates as reflection
from the inner accretion disc, the detection of correlated variation
of the ionised Fe\,K$\alpha$ emission line with the continuum implies that
the accretion disc at $\sim 100$r$_g$ has the same view of the central source 
as a distant observer, whereas the inner disc sees a much reduced variation
in illumination.

\item An alternative explanation for the lack of variation
is that the emission arises from a reflecting region that has a physically
large size ($\ga 1$ light-week).  The lack of Fe\,K$\alpha$ emission
may be achieved by scattering and absorption of nuclear emission
either by an extended zone of ionised gas with $\log\xi \sim 2-3$ or
by a very highly ionised zone ($\log\xi > 4$) accompanied by some
further absorption.  It is possible that this component could be
reflection from a hot dense wind or atmosphere such as that modelled
by \citet{proga04}.  Further modelling of such reflecting regions is
needed to test this explanation.

\item There appear to be discrete absorption lines at observed
energies 6.9 and 7.2\,keV.  The identification of the lines is not
secure, but they may arise in an outflow of velocity $\sim
13,000$\,km\,s$^{-1}$.  In the blurred reflection explanation, these
lines cannot arise from the same region as the reflection, as they
would be blurred themselves, so their appearance in the low state
of Mrk\,766 would need to be explained as a dependence of
opacity on the flux state of the source. 

\item The results from the PCA could alternatively be explained by 
models of ionised absorption in which the fraction of the source covered
is variable.  The variability timescale would imply an origin at $\sim 100$\,r$_g$
for the absorbing material. It is possible that the disc wind model
proposed by \citet{done} may be able to explain both the principal component
spectra and the spectral variability, if the wind is clumpy.

\end{itemize}

\begin{acknowledgements}
This paper is based on observations obtained with {{\it XMM-Newton}}, 
an ESA science mission with instruments and contributions 
directly funded by ESA Member States and NASA.  
TJT acknowledges funding by NASA grant NNG05GL03G.
We thank Kirpal Nandra and Stuart Sim for useful discussions.
\end{acknowledgements}


\begin{thebibliography}{99}
\setlength\itemsep{0mm}
\bibitem[Arnaud(1996)]{arnaud}
         Arnaud, K., 1996, 
Astronomical Data Analysis Software and Systems V, A.S.P. Conference Series, 
ed. G.H. Jacoby \& J. Barnes, 101, 17
\bibitem[Boller et al.(2001)]{boller}
       Boller, Th., Keil, R., Truemper, J., O'Brien, P.T., Reeves, J.N. \& Page, M.
       2001, A\&A, 365, L146
\bibitem[Boroson \& Green(1992)]{boroson}
       Boroson, T.A. \& Green, R.F. 1992, \apjs, 80, 109
\bibitem[Creshaw, Kraemer \& George(2003)]{crenshaw03}
        Crenshaw, D.M., Kraemer, S.B. \& George, I.M. 2003, ARA\&A,  41, 117 
\bibitem[Done et al.(2006)]{done}
        Done, C., Sobolewska, M.A., Gierlinski, M. \& Schurch, N.J. 2006,
       \mnras, in press, astro-ph/0610078
\bibitem[Fabian et al.(2002)]{fabian}
         Fabian, A.C., Vaughan, S., Nandra, K. et al. 2002, \mnras, 335, L1 
\bibitem[Ferland et al.(1992)]{ferland92}
         Ferland, G.J., Peterson, B.M., Horne, K., Welsh, W.F. \& Nahar, S.N. 1992, \apj, 387, 95
\bibitem[Francis et al.(1992)]{francis}
         Francis, P.J., Hewett, P.C., Foltz, C.B. \& Chaffee, F.H. 1992, \apj, 398, 476 
\bibitem[George \& Fabian(1991)]{georgefabian}
         George, I.M. \& Fabian, A.C. 1991, \mnras, 249, 352
\bibitem[Guilbert \& Rees(1988)]{guilbertrees88}
         Guilbert, P.W. \& Rees, M.J. 1988, \mnras, 233, 475 
\bibitem[Jansen et al.(2001)]{jansen}
  Jansen, F., Lumb, D., Altieri, B. et al. 2001, A\&A, 365, L1
\bibitem[Kallman et al.(2004)]{kallman04}
         Kallman, T., Palmeri, P., Bautista, M.A., Mendoza, C. \& Krolik, J.H.
         2004, \apjs, 155, 675
\bibitem[King \& Pounds(2003)]{kingpounds}
         King, A. \& Pounds, K.A. 2003, \mnras, 345, 657
\bibitem[Kotani et al.(2000)]{kotani}
         Kotani, T., Ebisawa, K., Dotani, T., Inoue, H., Nagase, F., Tanaka, Y. \& Ueda, Y. 2000,
         \apj, 539, 413
\bibitem[Kraemer et al.(2005)]{kraemer}
         Kraemer, S., George, I.M., Crenshaw, M. et al. 2005, \apj, 633, 693
\bibitem[Liedahl \& Torres(2005)]{liedahl}
         Liedahl, D.A. \& Torres, D.F. 2005, Canadian J. Phys., 83, 1177
\bibitem[Laor et al.(1991)]{laor91}
         Laor, A. 1991, \apj, 376, 90
\bibitem[Lightman \& White(1988)]{lightmanwhite88}
        Lightman, A.P. \& White,T.R. 1988, \apj, 335, 57
\bibitem[Magdziarz \& Zdziarski(1995)]{magdziarz95}
         Magdziarz, P. \& Zdziarski, A.A. 1995, \mnras, 273, 837
\bibitem[Markowitz et al.(2006)]{markowitz}
         Markowitz, A., Papadakis, I., Ar\'{e}valo, P., Turner, T.J., Miller, L.
          \& Reeves, J.N. 2006, \apj, in press, astro-ph/0611072
\bibitem[Matt et al.(1996)]{matt96}
         Matt, G., Fabian, A.C. \& Ross, R.R. 1996, \mnras, 278, 1111
\bibitem[Matt et al.(2000)]{matt00}
         Matt, G., Perola, G.C., Fiore, F., Guainazzi, M., Nicastro, F., 
         \& Piro, L. 2000, A\&A, 363, 863
\bibitem[Merloni et al.(2006)]{merloni06}
         Merloni, A., Malzac, J., Fabian, A.C. \& Ross, R.R. 2006, \mnras, 370, 1699
\bibitem[Miller et al.(2006a)]{miller06}
         Miller, L., Turner, T.J., Reeves, J.N. et al. 2006a, A\&A, 453, L13
\bibitem[Miller et al.(2006b)]{jmiller}
         Miller, J.M., Raymond, J., Homan, J. et al. 2006b, \apj, 646, 394
\bibitem[Miniutti et al.(2003)]{miniutti03}
         Miniutti, G., Fabian, A.C., Goyder, R. \& Lasenby, A.N. 2003, \mnras, 344, 22
\bibitem[Miniutti \& Fabian(2004)]{miniutti04}
         Miniutti, G. \& Fabian, A.C. 2004, \mnras, 349, 1435
\bibitem[Mittaz et al.(1990)]{mittaz}
         Mittaz, J.P.D., Penston, M.V. \& Snijders, M.A.J. 1990, \mnras, 242, 370
\bibitem[Nandra et al.(1997)]{nandraea97}
         Nandra, K., George, I.M., Mushotzky, R.F., Turner, T.J. \& Yaqoob, T. 
         1997, \apj, 477, 602 
\bibitem[Nayakshin \& Kazanas(2002)]{nayakshin}
         Nayakshin, S. \& Kazanas, D. 2002, \apj, 567, 85
\bibitem[Osterbrock \& Pogge(1985)]{osterbrock}
         Osterbrock, D.E. \& Pogge, R.W. 1985, \apj, 297, 166
\bibitem[Perola et al.(2002)]{perola02}
         Perola, G.C., Matt, G., Cappi, M., Fiore, F., Guainazzi, M., Maraschi, L., Petrucci, P.O. \& Piro, L. 2002, A\&A 389, 802  
\bibitem[Pounds et al.(2003a)]{pounds03a}
         Pounds, K., Reeves, J.N., Page, K.L., Wynn, G.A. \& O'Brien, P.T. 2003a,
         \mnras, 342, 1147
\bibitem[Pounds et al.(2003b)]{pounds03b}
        Pounds, K.A., Reeves, J.N., King, A.R., Page, K.L., O'Brien, P.T. \& Turner, M.J.L.
        2003b, \mnras, 345, 705
\bibitem[Pounds et al.(2004)]{pounds04}
        Pounds, K.A., Reeves, J.N., Page, K.L. \& O'Brien, P.T. 2004, \mnras, 605, 670
\bibitem[Poutanen(2001)]{poutanen}
        Poutanen, J., 2001, Advances in Space Research, 28, 267
\bibitem[Press et al.(1992)]{numrecipes}
         Press, W.H., Teukolsky, S.A., Vetterling, W.T. \&
          Flannery, B.P. 1992, ``Numerical Recipes'', second edition,
          (Cambridge University Press)
\bibitem[Proga \& Kallman(2004)]{proga04}
         Proga, D. \& Kallman, T. 2004, \apj, 616, 688
\bibitem[Reeves et al.(2003)]{reeves03}
         Reeves, J.N., O'Brien, P.T. \& Ward, M.J. 2003, \apj, 593, L65
\bibitem[Reeves et al.(2004)]{reeves04}
         Reeves, J.N., Nandra, K., George, I.M., Pounds, K.A., Turner, T.J. \&
         Yaqoob, T. 2004, \apj, 602, 648
\bibitem[Reynolds et al.(2004)]{reynolds} 
         Reynolds, C.S., Wilms, J., Begelman, M.C,
         Staubert, R. \& Kenziorra, E. 2004, \mnras, 349, 1153
\bibitem[Ross et al.(1996)]{ross}
         Ross, R.R., Fabian, A.C. \& Brand, W.N. 1996, \mnras, 278, 1082 
\bibitem[Ross \& Fabian(2005)]{rossfabian}
         Ross, R.R. \& Fabian, A.C. 2005, \mnras, 358, 211 
\bibitem[Sim(2005)]{sim05}
         Sim, S. 2005, \mnras, 356, 531
\bibitem[Smith et al.(2005)]{smith}
        Smith, J.E., Robinson, A., Young, S., 
        Axon, D.J. \& Corbett, E.A. 2005, \mnras, 359, 846
\bibitem[Str\"{u}der et al.(2001)]{struder}
  Str\"{u}der, L., Briel, U., Dennerl, K. et al. 2001, A\&A, 365, L18
\bibitem[Taylor et al.(2003)]{taylor}
         Taylor R.D., Uttley, P. \& McHardy, I.M. 2003, \mnras, 342, L31 
\bibitem[Tanaka et al.(1995)]{tanaka95} 
         Tanaka, Y., Nandra, K., Fabian, A. et al. 1995, Nature, 375, 659 
\bibitem[Turner et al.(2005)]{turnerea05}
         Turner, T.J., Kraemer, S.B., George, I.M., Reeves, J.N. \& Bottorff, M.C.
          2005, \apj, 618, 155
\bibitem[Turner et al.(2006)]{turner766}
         Turner, T.J., Miller, L., George, I.M. \& Reeves, J.N. 2006, A\&A, 445, 59
\bibitem[Uttley et al.(2004)]{uttley}
         Uttley, P., Taylor, R.D., McHardy, I.M. et al, 2004 \mnras, 347, 1345 
\bibitem[Vaughan \& Fabian(2003)]{vaughan03}
         Vaughan, S. \& Fabian, A.C. 2003, \mnras, 341, 496
\bibitem[Vaughan \& Fabian(2004)]{vaughanfabian04}
         Vaughan, S. \& Fabian, A.C. 2004, \mnras, 348, 1415 
\bibitem[Zdziarski et al.(1995)]{zdziarski95}
         Zdziarski, A.A., Johnson, W.N., Done, C., Smith, D., 
         \& McNaron-Brown, K. 1995, \apj 438, L63 
\bibitem[Zycki \& Czerny(1994)]{zycki94}
         \.{Z}ycki, P.T. \& Czerny, B. 1994, \mnras, 266, 653 
\end{thebibliography}
\end{document}